\documentclass[twocolumn,showpacs,preprintnumbers,amsmath,amssymb]{revtex4}
\usepackage{graphics, lscape, afterpage, longtable, bm}
\begin{document}
\title{Mode-Locking in Driven Disordered Systems as a Boundary-Value Problem}
\author{William Kung}
\affiliation{Department of Materials Science and Engineering, Northwestern University, Evanston, Illinois 60208}
\author{M. Cristina Marchetti}
\affiliation{Department of Physics, Syracuse University, Syracuse, New York 13244}
\date{\today}

\begin{abstract}
We study mode-locking in disordered media as a boundary-value problem.  Focusing on the simplest class of mode-locking models which consists of a single driven overdamped degree-of-freedom, we develop an analytical method to obtain the shape of the Arnol'd tongues in the regime of low ac-driving amplitude or high ac-driving frequency. The method is exact for a  scalloped pinning potential and easily adapted to other pinning potentials.  It is complementary to the analysis based on the well-known Shapiro's argument that holds in the perturbative regime of large driving amplitudes or low driving frequency, where the effect of pinning is weak. 
\end{abstract}


\maketitle

\section{Introduction}
The phenomenon of mode-locking is a general feature of nonlinear dynamical systems.  It consists
of the resonant response to an external periodic force that occurs when a characteristic frequency of the driven system matches or {\em locks onto} the driving frequency.  In the mode-locked region the system traces periodic orbits in phase space.  Outside the region of mode locking, the system may follow quasiperiodic orbits or march towards the onset of chaos. 

Examples of systems that exhibit mode-locking abound in nature.  In 1665, Huygens discovered the spontaneous synchronization of swinging pendulum clocks in close proximity to one another.  On the celestial scale, the moon's period of rotation locks onto its period of revolution about Earth in a 1:1-ratio, so it always presents the same face to Earth.  The rotation of Mercury is also locked onto the Sun such that there are three rotations for every two orbits, constituting a 3:2-mode locking. Mode-locking to external periodic stumuli is a common feature in biology.  Examples include the cell cycle in budding yeast~\cite{Siggia2005}, the swimming and heartbeat networks of medicinal leeches~\cite{Calabrese1992}, and the rhythmic behavior produced by neuronal networks~\cite{Coombes1999}.  Mode-locking is also relevant in such diverse phenomena as vortex shedding~\cite{Mureithi1998}, singing sand dunes~\cite{Andreotti2004}, and multimode lasers~\cite{Haus2000}.  In condensed-matter physics, Josephson junction arrays~\cite{Das1997}, driven superconducting vortices~\cite{Karapetrov2005, Besseling2004}, and charged-density waves (CDW)~\cite{Gruner1988, Fukuyama1978, Kolton2001, Fisher1985, Alstrom1987} provide convenient settings for studying mode-locking.  

Mathematicians have long used the language of ${\it{maps}}$ to describe the trajectories followed by dynamical systems in phase space and thus infer their physical properties.   In particular, circle maps~\cite{Jensen1984, Pradhan2002}, along with the mathematical tools of bifurcation theory and return map, lay the foundation for much of the theoretical modeling and analysis of mode-locking phenomena.  Conceptually, the simplest evolution equation that yields mode-locking consists of a single overdamped degree-of-freedom (DOF) $\phi$ in a periodic {\it{pinning}} potential $V_p(\phi)$ of stength $h$, driven by an external periodic force of frequency   $\omega$. The dynamics is described by the equation
\begin{eqnarray}
\frac{d\phi}{dt}&=&F_0+F_1\cos\omega t+F_p(\phi)\;,
\label{EOM}
\end{eqnarray}
where $F_p(\phi)=-h\frac{dV_p\left(\phi\right)}{d\phi}$  is the pinning force.  Mode-locking occurs when the time-averaged velocity locks on to a rational multiple of the external driving frequency,
\begin{eqnarray}
\left\langle\frac{d\phi}{dt}\right\rangle_{{\rm{cycle}}}\equiv&&\int_{2\pi/\omega}\frac{1}{2\pi/\omega}\frac{d\phi}{dt}\nonumber\\
=&& F_0+\left\langle F_p \right\rangle_{{\rm{cycle}}}\equiv\omega_d=\frac{p}{q}\,\omega
\label{wd1}
\end{eqnarray}
for a region of nonzero area in the $(F_0, F_1)$-space.  The regions of  the $(F_0, F_1)$ parameter space where mode-locking occurs are known as Arnol'd tongues~\cite{arnoldtongue}.  The time-averaged velocity exhibits a devil's-staircase (DS) behavior:  there exists a mode-locked plateau corresponding to each rational $p/q$.  While there have been much work on the fractal dimensionality of the set of gaps between the mode-locked steps in the staircase~\cite{Jensen1984, Halsey1986, Biham1989},  the calculation of the width of the mode-locked steps has mainly relied on an argument put forward by Shapiro for a single-particle model of  CDWs~\cite{Shapiro1964, Thorne1986}.  In this case the DOF  $\phi$ is  the phase of the CDW, while $F_0$ and $F_1$ are the amplitudes of the dc and the ac driving electric fields, respectively.  The time-averaged velocity corresponds to  the drift velocity $\omega_d$ of the CDW, which in turn determines  the CDW current.

Most available results on mode locking have been obtained numerically.  One of the rare analytical results is for a single DOF in a cosine pinning potential, $V_p(\phi)=h\cos\phi$.   In this case the Arnol'd tongues are symmetric with respect to the mirror axis centered at the apex of the tongue and pinch to zero width for all parameters in its phase space, as shown in Fig.~\ref{Fig0}.  On the other hand, numerics have shown that in extended systems consisting of many coupled DOFs the Arnol'd tongues are generally asymmetric and never pinch to zero width. In this work, we reexamine single DOF model for various pinning potentials and show that the cosine pinning  is special. For a generic pining potential the tongues are asymmetric and do not
pinch to zero as soon as the pinning potential contains any higher harmonics.  This more complex shape of the Arnol'd tongues may be the result of the collective behavior of many coupled degrees of freedom, but it arises even for a single particle, provided the pinning potential differs from a simple cosine. Of course in an extended system collective effects renormalize the pinning potential, so that even a bare cosine pinning yields asymmetric Arnol'd tongues. We show below that for generic pinning potential the Shapiro's method always fails at small $F_1$, where the asymmetry of the Arnol'd tongue is most apparent. For the specific case of scalloped potential, we develop an exact analytical mehtod for the calculation of the mode-locking behavior of one DOF for $F_1$ small enough  that the single DOF does not hop from one scallop to the next.  This is precisely the regime  where the Shapiro's argument always fails. Our  analytical solution fro small $F_1$ also provides dynamical constraints that are both necessary and sufficient for the full determination of the mode-locked step widths.

This work  serves as a starting point in our effort  to construct a mean-field theory for the general phenomenon of mode-locking in extended media that is amenable to analytical analysis in useful limits.  To do so, we must first test analytical approximations at the single-particle level and identify a simple, yet generic, pinning potential as our prototypical starting model.  To achieve these two first-step goals, we study in detail the case of 1:1-mode locking and consider two specific  periodic pinning potential $G[\phi]$, a scalloped parabolic and an impure cosine potential.  In section II we review Shapiro's method for calculating the width of the mode-locked steps.  We  then consider the one-particle mode-locking dynamics in the scalloped parabolic pinning potential in section III.  This form of potential has the advantage of permitting exact analytical solution in the no-hopping regime.  In section IV, we  repeat our analysis for the impure cosine pinning potential.  Section V concludes our paper.

\section{Shapiro's method in Calculating Mode-Locked Step Widths}
 
Adapting an argument proposed  by Shapiro~\cite{Shapiro1964}, Thorne {\it{et. al.}}  computed the width of the harmonic and subharmonic mode-locked steps for the one-DOF model of CDW given in Eq. (\ref{EOM})~\cite{Thorne1986}.  While their calculation explains the occurrence of mode-locking in terms of the lowering of the pinning energy, their results do not agree with numerical results for the shape of the Arnol'd tongues in the regime of low ac-driving amplitude or high ac-driving frequency.  This regime corresponds to the case where the particle dynamics is confined to one period of the  pinning potential. To show how Shapiro's method fails  in this regime, we first review the calculation of Thorne et al.~\cite{Thorne1986}.  

\begin{figure}[t]
\vspace{2mm}
\hspace{23mm}
\includegraphics{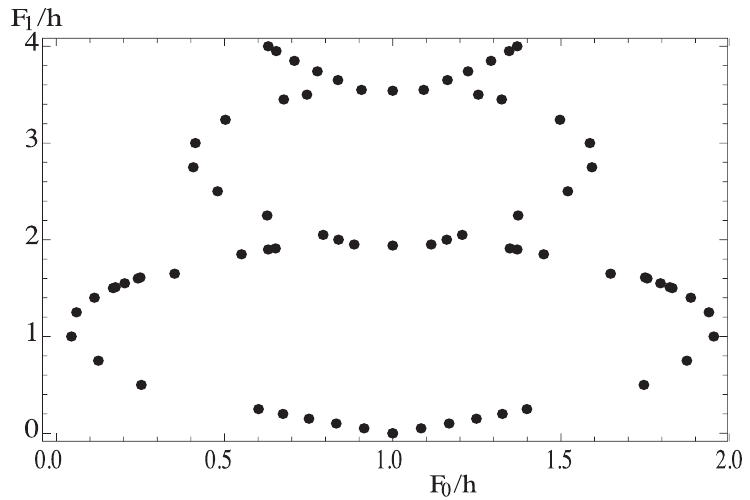}
\vspace{2mm}
\caption{1:1 Arnold tongue for pure cosine pinning potential.}
\label{Fig0}
\end{figure}

It is instructive to first consider the single DOF model of Eq.~(\ref{EOM}) for  a cosine pinning potential.
We note that in the absence of pinning the equation of motion has the exact solution 
$\phi(t)=F_0t+\frac{F_1}{\omega}\sin\omega t +\phi_0$, with $\phi_0$ a constant, which gives the trivial result
$\omega_d=F_0$. In the presence of pinning we let 
\begin{equation}\label{soln}
\phi(t)=\omega_dt+\frac{F_1}{\omega}\sin\omega t +\phi_0(t)\;,
\end{equation}
 with $\omega_d$ and $\phi_0(t)$ to be determined. We note that Eq.~(\ref{EOM}) contains two characteristic frequency scales: the frequency $\omega$ of the external drive and the frequency $h$ of temporal variations of the phase due to the pinning potential. In the limit of large drive,  the particle moves rapidly over the pinning potential  and the temporal variations due to pinning are small. Following Ref.~\cite{Thorne1986} we then look for a solution where $\phi_o\approx{\rm constant}$, independent of time. We expect this approximation will apply for large $F_1$ and weak pinning. It turns out to be essentially exact for a simple
 cosine pinning potential, but it
 fails for small values of $F_1$ for arbitrary periodic pinning.  Substituting Eq.~(\ref{soln}) into Eq.~(\ref{EOM}), 
 utilizing the Bessel function summation formula  
\begin{eqnarray}
\exp\left[-\imath\left(\frac{F_1}{\omega}\right)\sin\,\omega t\right]=\sum_{p=-\infty}^{\infty}J_p\left(\frac{F_1}{\omega}\right)\,\exp\left(-\imath p\omega t\right),\nonumber\\
\end{eqnarray}
and averaging over a cycle, we obtain
\begin{eqnarray}\label{modelocked}
&&\omega_d=F_0+\left\langle F_p \right\rangle_{{\rm{cycle}}}\;,\\
&&\label{Favg1}\left\langle F_p \right\rangle_{{\rm{cycle}}}=h\sin\phi_0\sum_{p=1\atop (\omega_d=p\omega)}^{\infty} (-1)^p J_p(F_1/\omega)\;,
\end{eqnarray}
where the sum is over all $p$ such that $\omega_d=p\omega$. Similarly the mean pinning energy is given by 
\begin{equation}
\left\langle V_p \right\rangle_{{\rm{cycle}}}=h\cos\phi_0\sum_{p=1\atop (\omega_d=p\omega)}^{\infty} (-1)^p J_p(F_1/\omega)\;.
\end{equation}
A mode-locked state occurs when  $\omega_d$ in Eq.~(\ref{modelocked}) can be kept constant (and equal to $p\omega$) for a range of values
of $F_0$ and $F_1$ by adjusting the phase $\phi_0$. For a cosine pinning potential the range of $\phi_0$ that renders $\omega_d={\rm constant}$ is identical to the range of $\phi_0$ where
$\left\langle V_p \right\rangle_{{\rm{cycle}}}<0$. In this case the region of parameters where the system is mode-locked coincides with the region where the energy  of the mode-locked state is lower than that of the unlocked state. This is not, however, the case for arbitrary pinning potential, as we will see below. 
In general the condition that energy of the mode-locked state be lower than that of the unlocked state is 
necessary for their stability of the mode-locked state, but not sufficient (see Fig. \ref{impure} below).  For pure cosine pinning case only harmonic steps are obtained. The $p/1$ step   occurs for a range
$(\Delta F_0)_{p/1}$ of values of $F_0$ given by $(\Delta F_0)_{p/1}=2h|J_p(F_1/\omega)|$. The 
Arnol'd tongues for this  case are shown in Fig.~\ref{Fig0}.

Thorne and collaborators adapted this argument to an arbitrary periodic pinning potential that
can be  expressed as a Fourier sum: 
\begin{eqnarray}
V_p(\phi)&=&\frac{a_0}{2}+\sum_{q=1}^{\infty}\,a_q\,\cos\,q \phi\;,
\end{eqnarray}
Proceeding as in the single cosine case, using again the Bessel summation formula and averaging over a cycle, we find that $\omega_d$ is determined by Eq.~(\ref{modelocked}), where the
 time-averaged pinning force is zero unless  $\omega_d=p\omega/q$, for which
\begin{eqnarray}
%\left\langle F_p\right\rangle\rightarrow
\left\langle F_p\right\rangle_{cycle}&=&2\sum_{q=1}^{\infty}\sum_{p=-\infty}^{\infty}qa_qJ_p\left(\frac{qF_1}{\omega}\right)\sin\,q\phi_0.\nonumber\\
\label{arnold}
\end{eqnarray}
Thus, for a given $F_1$, the constant phase $\phi_0$ can be adjusted to compensate changes in $F_0$ such that the time-average phase velocity $\omega_d$ stays locked to rational values of the external driving frequency.  Furthermore, Eq. (\ref{arnold}) suggests that the width of the Shapiro steps oscillates with  $F_1$, which has indeed been observed in experiments~\cite{Besseling2004, Lubbig1974}.   
These authors further assumed that the range of parameters where the system is mode-locked
coincides with those where the mean pinning energy in the mode locked state is lower than its value 
$a_0/2$ in the unlocked state, i.e., 
%
%\begin{widetext}
\begin{eqnarray}
\delta \langle V_p\rangle_{cycle}&=&\sum_{q=1}^{\infty}\sum_{p=-\infty}^{\infty}\,a_qJ_p\left(\frac{qF_1}{\omega}\right)\cos\left(q\phi_0\right)<0\;.\nonumber\\
\label{thorne}
\end{eqnarray}
%\end{widetext}
%
The mean pinning force given by Eq.~(\ref{arnold})
cancels the DC drive $F_0$ in a range $-\phi_p(F_0,F_1)\leq\phi_0\leq \phi_p(F_0,F_1)$. The condition that $\phi_p(F_0,F_1)\in [0,2\pi]$ can then be used to determine the boundaries of the mode-locked
regions in the $(F_0,F_1)$ plane. Thorne et al. use the condition  $\delta \langle V_p\rangle_{cycle}<0$ to determine the Arnold tongues. The latter holds for a range 
$-\phi_m(F_0,F_1)\leq\phi_0\leq \phi_m(F_0,F_1)$. 
For pure cosine pinning $\phi_m(F_0,F_1)=\phi_p(F_0,F_1)$. For arbitrary pinning potentials
$\phi_m(F_0,F_1)\geq\phi_p(F_0,F_1)$, resulting in an overestimate of the mode-locked regions.
Regardless of the condition used for mode-locking, a consequence of Eqs. (\ref{arnold}) and (\ref{thorne}) is that each Arnol'd tongue is symmetric about a central axis
for any periodic pinning potential. In the case of 1:1-mode locking, the tongue is symmetric about the axis $F_0=1$,  since $J_p(0)=0$ for all $p\neq 0$ and $J_0(0)=1$ assuming appropriate normalization.  For the 1:1-mode locking case, we will show explicitly, via exact numerical calculations, that this symmetry of Arnol'd tongue is violated in the regime where the influence of ac-drive is large, {\it{i.\,e.}} when the drive amplitude $F_1$ is small or the drive frequency $\omega$ is small ({\it{c.f.}} Fig.~\ref{scap}), for two specific cases of the periodic pinning potential corresponding to the scalloped parabolic and the impure cosine pinning.

\begin{figure}[t]
\vspace{2mm}
\hspace{23mm}
\includegraphics{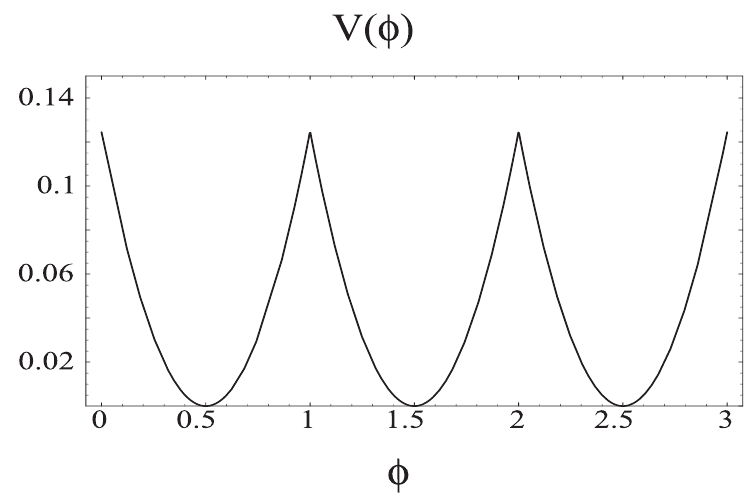}
\vspace{2mm}
\caption{The scalloped parabolic pinning potential, defined by Eq. (\ref{scalloped1}).}
\label{Fig1}
\end{figure}
\section{Scalloped Parabolic Pinning Potential}

\begin{figure}[t]
\vspace{2mm}
\hspace{23mm}
\includegraphics{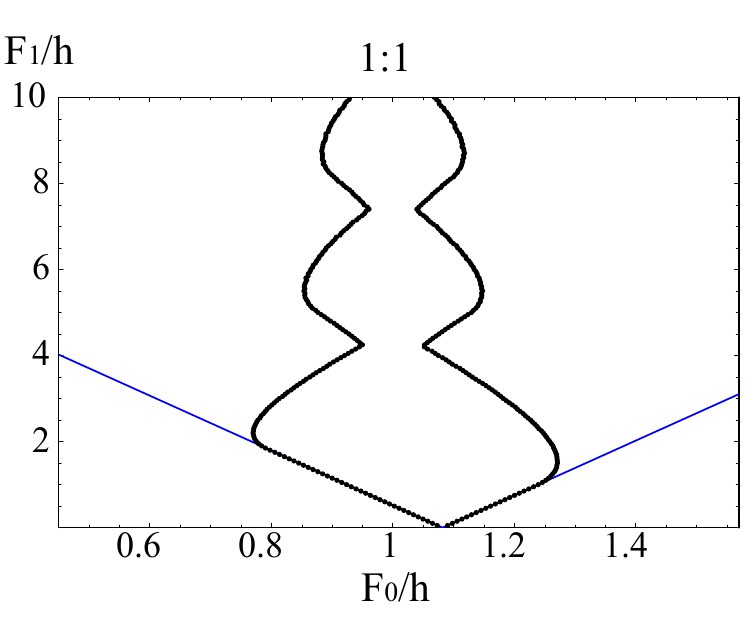}
\vspace{2mm}
\caption{1:1 Arnold tongue for scalloped parabolic pinning potential:  Thorne's method versus "exact" method for small $F_1$}
\label{tongueexact}
\end{figure}

\begin{figure}[t]
\vspace{2mm}
\hspace{23mm}
\includegraphics{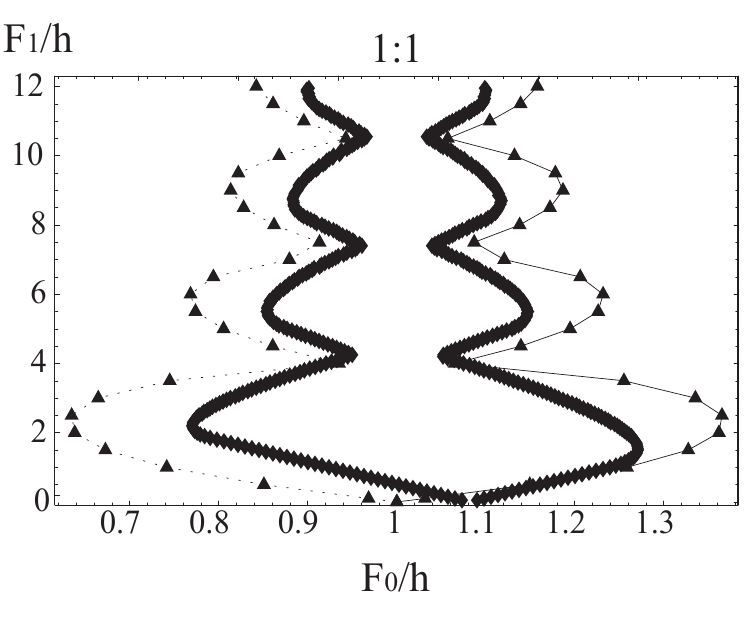}
\vspace{2mm}
\caption{Plot of Arnol'd tongue for 1:1 mode-locking step with the scalloped parabolic pinning potential.  Black triangles represent data points obtained from the zeros of the pinning energy using Thorne's method.  Numerically, the first 100 terms of the sum were included in our summation over the Bessel functions.  The black diamonds represent data from numerical results.}
\label{scap}
\end{figure}

We first consider the case of scalloped parabolic pinning, whose potential can be expressed either in terms of the floor function,  $\lfloor x\rfloor$, defined as the largest integer less than or equal to its argument $x$, or as a Fourier series:
\begin{eqnarray}
\label{scalloped1}
V_{\rm{scalloped}}(\phi)&=&\frac{1}{2}\left(\phi-\lfloor \phi\rfloor-\frac12\right)^2\\
&=&\frac{1}{24}+\sum_{q=0}^{\infty} \frac{1}{2\pi^2q^2}\cos 2\pi q \phi\;.
\label{scalloped}
\end{eqnarray}
A plot of the scalloped parabolic potential is shown in Fig.~\ref{Fig1}.  
In this case the double sum in the expression for the time-averaged pinning force, Eq. (\ref{arnold}), reduces to a single infinite sum over terms for which $p=q$.  Using Eq. (\ref{scalloped}), the cycle-averaged pinning force is given by:
\begin{eqnarray}
\langle F_p(\phi_0)\rangle_{cycle}&=&\sum_{q=0}^{\infty} \frac{1}{\pi q}\,J_q(qF_1)\,\sin 2\pi q \phi_0\;.
\end{eqnarray}
For the scalloped pinning potential of Fig.~\ref{Fig1} the  pinning force is piece-wise linear and the dynamics of the driven particle can be solved exactly within each period of the pinning potential. This solution is expected to be exact for small values of $F_1$, provided the particle does not hop from one scallop to the next i the time scale of the external drive. Within one period Eq. ({\ref{EOM}}) is simply
\begin{eqnarray}\label{EOMscallop}
\frac{d\phi}{dt}+\phi&=&\left(F_0+\frac 12\right)+F_1\cos\,2\pi t\;.
\end{eqnarray}
For simplicity we discuss in detail only mode-locked steps with $p=1$. In this case Eq.~(\ref{EOMscallop}) must be solved with the boundary conditions
\begin{eqnarray}
\label{bc1}
\phi(t_J+nT)&=&0\;,\\
\phi(t_J+(n+1)T)&=&1\;,
\label{bc2}
\end{eqnarray}
where $T=2\pi q/\omega$.  The jump time $t_J$ plays the role of the constant phase $\phi_0$ in the previous section. Equation (\ref{bc1}) determines the constant of integration for our first-order ODE, while Eq. (\ref{bc2}) yields the relationship between $F_0$ and $F_1$ corresponding to mode-locking.  This can be rewritten as
\begin{eqnarray}
\frac{1}{1-e^{-1}}=\left[F_0+\frac 12 +\frac{F_1}{h^2+\omega^2}(\cos\,2\pi t_j-2\pi\sin\,2\pi t_j)\right]\;.\nonumber\\
\label{constraint}
\end{eqnarray}
The condition that the values of $t_J$ in Eq.~(\ref{constraint}) must be in $[0,T]$ determines the boundaries $F_1(F_0)$ of the Arnol'd tongues, given by
\begin{eqnarray}
F_1\geq\pm \sqrt{1+(2\pi)^2}\left(\frac{1}{1-e^{-1}}-F_0-\frac 12\right)\;.
\label{dis}
\end{eqnarray}
These are the straight lines shown in Fig.~\ref{tongueexact}, along with the complete Arnol'd tongues for the 1:1 step obtained by exact numerics.  Clearly the analytical solution yields the exact value of $F_0$ for the onset of mode-locking at $F_1=0$ and also does an excellent job of fitting the exact Arnol'd tongues for small $F_1$, where the driven particle remains within a single scallop.
In contrast, Shapiro's argument fails most severely precisely in this region of small $F_1$, as
apparent from Fig.~\ref{scap} where the Arnol'd tongues obtained for the 1:1 mode-locked step by the Shapiro argument (triangles) are compared to the exact solution (diamonds).
 As mentioned in section II, Shapiro's method predicts symmetrical oscillation about the axis $F_0=1$.  However,  exact numerical solution reveals that the 1:1-mode locked step, in fact, originates from $F_0\approx1.1$.
This difference arises because the Shapiro method is essentially a perturbation theory about the high velocity state and becomes exact at large drives and very weak disorder. On the other hand for fixed driving frequency $\omega$ and small ac-drive amplitude $F_1$, or conversely  fixed $F_1$ and high frequency the driving force has only a weak effect and pinning dominates.  In this region the  time-averaged velocity is well approximated by  the instantaneous velocity.  This fact underlies our approximation scheme in which we solve exactly for the instantaneous dynamical solution for the CDW phase $\phi(t)$, assuming that the particle does not hop over any one scallop in the periodic pinning potential.  Neglecting the effect of scallop-hopping simplifies the dynamics considerably, as the elimination of the associated nonlinearity permits analytically tractable solutions. This method is complementary to the one developed by Shapiro.
Our method can be readily generalized to other harmonic and subharmonic mode-locking steps that involve no hopping between scallops and thus satisfy the constraint $p=1$.

\section{Impure cosine pinning potential}

For a cosine pinning potential the width of mode-locked steps pinches to zero periodically at finite values of the ac-driving amplitude, $F_1$.   This is well explained by Shapiro's argument~\cite{Thorne1986} based on its expression of the time-averaged pinning potential:  since the time-averaged pinning force has only one term for any particular values of $p$ and $q$ and it is a consequence of   the fact that pinning potential  contains only a single harmonic in its Fourier series.  To see this point explicitly, we  compare the two cases  of a pure cosine pinning potential function and of an {\it{impure}} cosine,  consisting of a sum of two different harmonics:
\begin{eqnarray}
V_{\rm{cosine}}(\phi)&=&\frac{1}{2\pi}\cos 2\pi \phi\;,\\
V_{\rm{impure}}(\phi)&=&\frac{1}{2\pi}\cos 2\pi \phi+\frac{0.1}{4\pi}\,\cos 4\pi \phi\;.
\end{eqnarray}
\begin{figure}[t]
\vspace{2mm}
\hspace{23mm}
\includegraphics{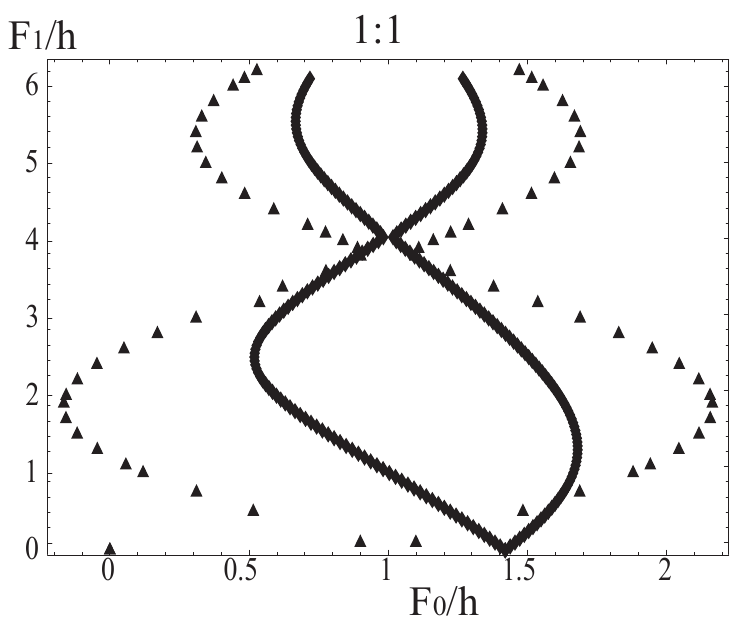}
\vspace{2mm}
\caption{Plot of the impure cosine pinning potential consisting of a sum of two harmonics.  Gray triangles represent data points obtained from the zeros of the pinning energy using Thorne's method.  The black diamonds represent data from numerics.}
\label{impure}
\end{figure}
Again we are focusing on 1:1-mode locked steps. For the impure cosine pinning, the  time-averaged pinning force is 
\begin{eqnarray}
\langle F_p(\phi_0)\rangle&\sim&J_1(F_1)\sin 2\pi \phi_0+0.1J_2\left(2F_1\right)\sin 4\pi\phi_0\nonumber\\
\end{eqnarray}
It is apparent from Fig.~\ref{impure}  that the addition of a small harmonic is sufficient to give a finite width to the tongues for all nonzero values of $F_1$.  Again Shapiro's method clearly fails at small $F_1$.  To analyze the dynamics in this region, we assume that there is no particle-hopping 
and use the analytical method discussed in the previous section.
To do this we fit  the impure cosine pinning potential  to a parabola over one cycle.  Using Mathematica, we find that the fitted pinning potential yields $V_{\rm{cosine}}^{F}(\phi)=0.186-1.17\phi+1.19\phi^2$, or equivalently, a scalloped parabolic pinning potential with a pinning strength of $h\approx 2.39$ over the same cycle.  The resulting comparison plot of the numerically obtained Arnol'd tongue for 1:1-mode locking and the ``V"-shaped asymptotes obtained analytically  in the no-hopping approximation is shown in Fig. (\ref{impurearnold}).  Again, there is remarkable agreement between our analytical approximation and the exact numerical solution in the small-drive regime. 
\begin{figure}[t]
\vspace{2mm}
\hspace{23mm}
\includegraphics{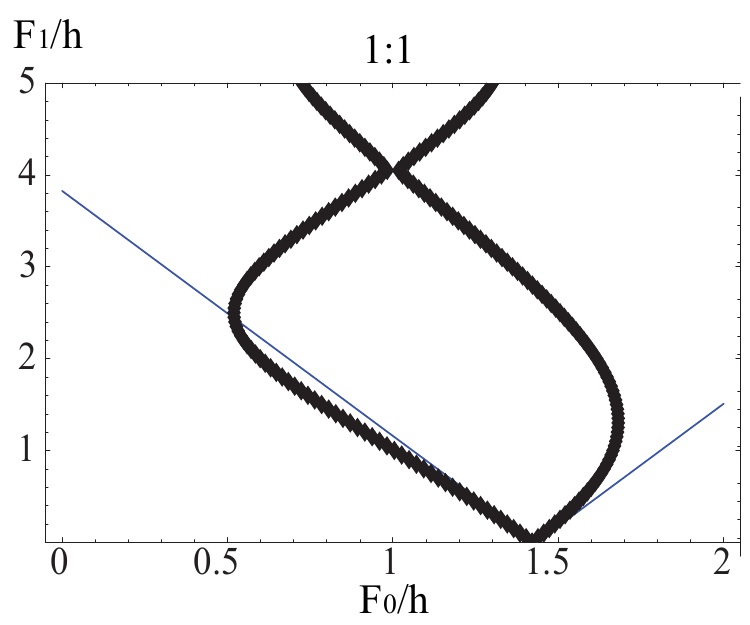}
\vspace{2mm}
\caption{Shape of the Arnol'd tongue for the 1:1 mode-locking step for an impure cosine pinning potential. The diamonds are exact numerical results and the V-shaped straight lines are the analytical approximation in the no-hopping regime, corresponding to the limit of small drive/high frequency.}
\label{impurearnold}
\end{figure}

\begin{figure}[t]
\vspace{2mm}
\hspace{23mm}
\includegraphics{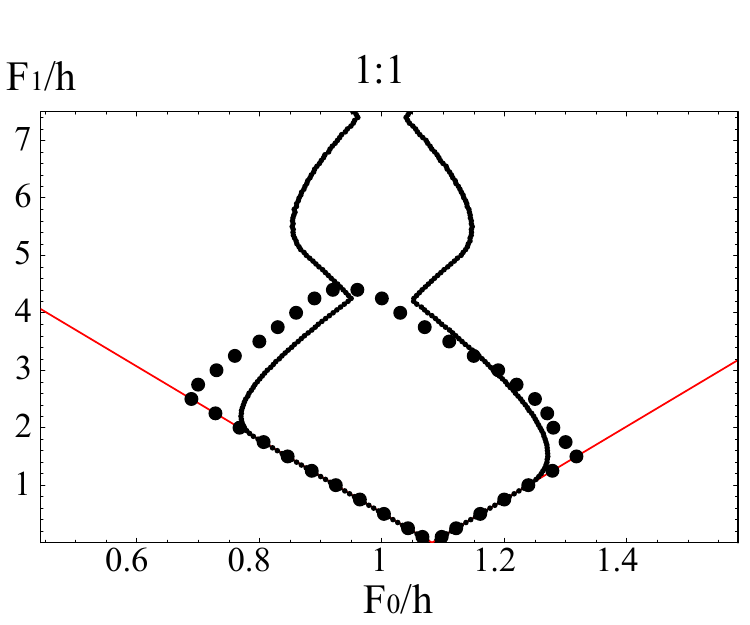}
\vspace{2mm}
\caption{The shape  of the Arnol'd tongue for the 1:1 mode-locking step for a scalloped parabolic pinning potential. The solid line are exact numerical results and the (red online) V-shaped straight lines correspond to the analytical approximation in the no hopping regime.  The region bounded by the black circles are points of $(F_0,F_1)$ which are explicitly checked to satisfy the constraint of $\left\vert \phi(t)\right\vert\leq 1$.}
\label{gr5}
\end{figure}

To further clarify the role of the no-hopping approximation used in our analytical calculation, we note that in obtaining Eq. (\ref{dis}), we implicitly assume that $\left\vert \phi(t)\right\vert\leq 1$, for $t\in[0,2\pi]$.   Self-consistency of our solution is then imposed by looking for the set of points in the region defined by Eq. (\ref{dis}) that explicitly satisfy $\left\vert \phi(t)\right\vert\leq 1$ for $t\in[0,2\pi]$.  As shown in Fig.~\ref{gr5}, the region of $(F_0,F_1)$, bounded by black circles, which explicitly satisfies the constraint $\left\vert \phi(t)\right\vert\leq 1$ is an excellent approximation to the lopsided pear-shaped region of the Arnol'd-tongue for $F_1<4$.  Beyond this region, the particle presumably hops over more than one scallop per drive period, and our analytical approximation would no longer apply.   

\section{Conclusion}

In summary, we have developed a simple analytical method to obtain the shape  of the Arnol'd tongues  in the regime of small ac-driving amplitude $F_1$ or high-driving frequency, where the driven particle does not hop between different periods of the driving potential within one period of the drive. This method is complementary to the perturbative one based on  Shapiro's argument~\cite{Shapiro1964, Thorne1986} that applies in the large $F_1$ or low frequency regime.  The method is exact for a scalloped pinning potential and is easily adapted to other pinning potentials by a simple fit.     Our method is easily adapted to the analysis of $p/q$ mode-locking  steps for arbitrary $q$ and $p=1$. 

As mentioned in the Introduction, the motivation of this work was to develop simple methods for the analysis of mode-locking steps that will serve as the starting point for the development of a mean-field theory of mode-locking in  systems composed of many interacting many degrees-of-freedom.  We hope to do this by combining the no-hopping approximation for the analysis of the low drive regime with the Shapiro argument for the study of the high-drive region.  Finally, while the phenomenon of mode-locking is pervasive in dynamical systems, each subfield has developed its own set of theoretical tools, often inspired and required by real-world applications.  Despite past attempts~\cite{MacKay1994}, a consensus is still lacking in regard to a unifying formalism in the description of mode-locking.   It is our further hope that our results would illuminate aspects of mode-locking, which are still not well understood and which would merit further study. 

\begin{acknowledgments}

We would like to thank Alan Middleton for helpful discussions.  This work was supported by NSF Grants  DMR-0305497 and DMR-0705105.

\end{acknowledgments}

\end{document}